\documentclass[prb]{revtex4}

\usepackage{graphicx}
\usepackage{dcolumn}
\usepackage{bm}

\begin{document}


\title{Low-energy properties of two-dimensional 
quantum triangular antiferromagnets: Non-perturbative
renormalization group approach}

\author{Satoshi Fujimoto}
 \affiliation{Department of Physics, Kyoto University, Kyoto 606-8502, Japan.}

\date{\today}

\begin{abstract}
We explore low temperature properties of quantum triangular Heisenberg 
antiferromagnets in two dimension in the vicinity of
the quantum phase transition at zero temperature.
Using 
the effective field theory described by
the $O(3)\times O(2)$ matrix Ginzburg-Landau-Wilson model
and the non-perturbative renormalization group
method, we clarify how quantum and thermal fluctuations
affect long-wavelength behaviors in the parameter region where
the systems exhibit a fluctuation-driven first order transition
to a long-range ordered state.
We show that at finite temperatures
the crossover from a quantum $\phi^6$ theory
to a renormalized two-dimensional classical nonlinear sigma model
region appears, and
in this crossover region, massless fluctuation modes
with linear dispersion {\it a la} spin waves govern
low-energy physics. 
Our results are partly in good agreement with the recent experimental
observations for the two-dimensional
triangular Heisenberg spin system, NiGa$_2$S$_4$.
\end{abstract}

\pacs{Valid PACS appear here}
\maketitle

\section{Introduction}

It has been argued for decades that geometrical frustration
gives rise to unusual magnetic properties in quantum antiferromagnets,
and may bring about an exotic
ground state such as a spin liquid
which is characterized by the absence of any type of spontaneous
symmetry breaking including magnetic long-range order (LRO), and
dimerization, etc.\cite{frust,vi,yos,kap,tou,vi2,and}
The concept of the spin liquid was first proposed by Anderson
in connection with the ground state of two-dimensional (2D) Heisenberg 
antiferromagnets (HAF) on a triangular lattice.\cite{and}
While extensive studies on frustrated magnets have revealed 
that the triangular HAF may exhibit the magnetic 
LRO,\cite{kami,kawa,kawa1,hk,lh,aza2,bha,mai,loi2,win,ant}
the Anderson's idea is still attracting much interest, and  
has been tested for its possible realization in other geometrically frustrated
systems such as pyrochlore and Kagome 
HAF, and multiple ring exchange spin 
models.\cite{cha,sac,rit,zeng,lech,mila,py,iso,moe,yam,koga,tsune,fuji,fuji2,ra,mis2,lim} 

Recently, Shimizu et al. reported that 
in an organic Mott insulator with spin $s=1/2$
on a triangular lattice, $\kappa$-(ET)$_2$Cu$_2$(CN)$_3$, 
no magnetic LRO is observed down to 32 mK.\cite{kano}
Their experimental results suggest the possibility of a new kind of 
a ground state including the spin liquid.
Subsequently, Nakatsuji et al. found that 
in a quasi-2D quantum triangular HAF with $s=1$,
NiGa$_2$S$_4$, there is no sign of LRO down to 0.35 K,
in spite of the existence of strong anferromagnetic interactions.\cite{naka} 
In the latter system, the specific heat coefficient $C_v$ shows 
the quadratic-temperature dependence $C_v\sim T^2$ at sufficiently
low temperatures, and the uniform spin susceptibility
is constant in the low temperature regions, indicating
the existence of low-lying massless excitations.
The origin of these unexpected low-temperature behaviors
has not yet been explained.

\begin{figure}
\includegraphics*[width=6cm]{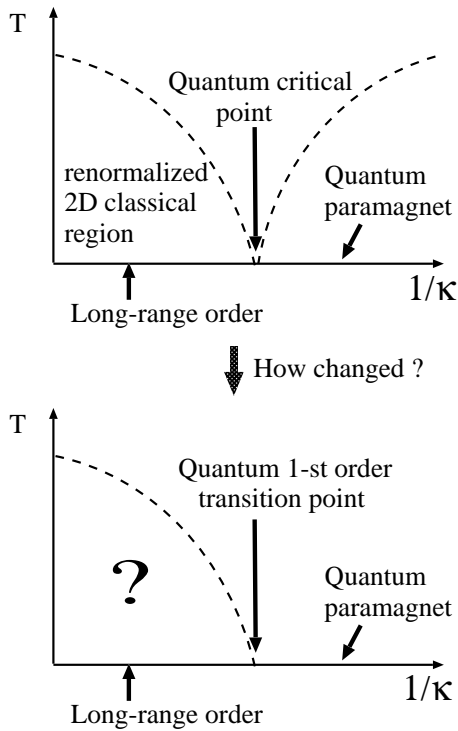}
\caption{\label{fig:schpha} Phase diagrams of quantum second order
phase transition (top) and quantum first order transition (bottom).
The vertical and horizontal axes are, respectively, 
temperature $T$ and a parameter $1/\kappa$ 
which controls quantum fluctuations.} 
\end{figure}

On the other hand, from theoretical point of view, 
there have been only a few works on
low-energy properties in the vicinity of
the quantum phase transition at $T=0$ in the 2D triangular 
HAF.\cite{aza1,chu} 
To understand the above-mentioned experimental observations
for quantum triangular HAF precisely, we need to develop
a theory which describes quantum  critical phenomena in these systems
at finite temperatures.
In contrast to the quantum case,
the three-dimensional (3D) classical stacked Heisenberg model,
which is equivalent to the 2D quantum model at $T=0$, has been
extensively studied by many 
authors.\cite{kami,kawa,kawa1,bha,mai,loi2,win,ant,zum2,zum,loi,ita,tis,del,pel,cal,cal2}
Even for the classical systems, the elucidation of the nature of
the phase transition has not yet been completed. 
Several theoretical works done by Zumbach, Loison and Schotte, 
Itakura, and Delamotte et al. indicate 
that the 3D classical triangular HAF show
a fluctuation-driven first order phase
transition.\cite{zum2,zum,loi,ita,tis,del}  
On the other hand, the loop expansion calculations carried out by
Pelissetto et al. and Calabrese et al. support the existence of the continuous
phase transition in these systems.\cite{pel,cal2} Also, most of experiments
seem to be in accordance with the continuous transition.\cite{plak}
However, the former point of view is quite intriguing, since
it implies that 
the phase transition of the 2D quantum version of these 
systems at $T=0$ may be
the (quantum) fluctuation-driven first order type.
Although the quantum second order phase transition and
the related quantum critical phenomena 
have been comprehensively explored so far,\cite{chn,hert,you2,aza1,chu,fis} 
long-wavelength properties which emerge
near the quantum fluctuation-driven first order transition
 have not been clarified sufficiently.
It is well-known that
in the case of the continuous quantum phase transition, 
critical phenomena just above the transition point at $T=0$
are described by a renormalized 2D classical theory.
In contrast, it is highly non-trivial how 
quantum and thermal fluctuations which induce a first order
 transition at zero temperature 
affect low-energy behaviors (See Fig.~\ref{fig:schpha}).
In this paper, we would like to address this issue 
for the 2D quantum triangular HAF.

Generally, in 2D quantum critical phenomena, the crossover
from the 2+1D quantum behaviors to the renormalized 2D classical ones occurs
at finite temperatures.
For the precise description of the quantum-classical crossover,
a promising theoretical approach may be
the non-perturbative renormalization group (RG) method.
This technique has been applied to classical frustrated magnets by
Delamotte et al. yielding fruitful results.\cite{del}
They 
showed that the non-perturbative RG method
successfully reproduces the RG equations for 
both the four-dimensional (4D) Ginzuburg-Landau-Wilson
(GLW) model and the 2D nonlinear sigma model, which are
the effective field theories for the 4D and 2D classical triangular HAF, 
respectively, and is expected to capture correct low-energy physics of
the 3D triangular HAF.
Thus, the non-perturbative RG method may be suitable for 
the investigation of
the dimensional (or quantum-to-classical) crossover phenomena
for these systems.
We utilize this remarkable merit of the approach to shed light on 
how quantum and thermal fluctuations control low-energy properties
in the vicinity of quantum fluctuation-driven first order transitions.

Our main results are as follows.
At finite temperatures,
the crossover from a quantum $\phi^6$ model to a renormalized 2D classical
system appears, and as $T$ decreases, 
long-wavelength behaviors are almost governed
by the $\phi^6$-fluctuations which eventually 
bring about the first order transition.
Also, it is found that in this crossover region, thermodynamic properties
are effectively determined by gapless excitations {\it a la} spin waves 
with a linear dispersion.
The presence of these low-lying excitations 
explains partly the experimental observations for NiGa$_2$S$_4$;
i.e. the quadratic $T$-dependence of the specific heat
coefficient, and the finite $T$-independent uniform spin susceptibility
at low temperatures.\cite{naka}

The organization of this paper is as follows.
In Sec.II, the effective field theory and the formulation
of the non-perturbative RG method are given.
In Sec.III, we present results on the RG flows
obtained by solving numerically the RG equations, which
demonstrate the existence of strong fluctuations driving the phase
transition at $T=0$ to the first order type.
In Sec.IV, we show that in the crossover region,
quasi-Gaussian fluctuations dominate low-energy properties, and
derive low-$T$ behaviors of 
the specific heat coefficient and the uniform spin susceptibility.
Summary of our results and discussion on the implication 
for the recent experimental observations for NiGa$_2$S$_4$ are given in Sec.V.

\section{Effective field theory and non-perturbative renormalization
group method}

\subsection{Quantum $O(3)\times O(2)$ 
matrix Ginzburg-Landau-Wilson model
for frustrated magnets}

Quantum phase transitions in two dimension which occur at $T=0$
can not be described by the usual $\phi^4$-type Ginzburg-Landau scheme, since
the long-range ordered phase exists only at $T=0$, and 
the order parameter becomes nonzero abruptly just at $T=0$.
In the case of 2D quantum non-frustrated HAF,
the low-energy properties 
of the quantum phase transition are successfully
explained in terms of the $O(3)$ nonlinear sigma model.\cite{chn}
In the derivation of the nonlinear sigma model from
the HAF, one merely postulates that 
the correlation length is larger than the lattice constant.
Thus the model describes long-wavelength physics of
both the ordered phase at $T=0$ and  
the disordered phase above the transition temperature, which 
are mainly governed by transverse fluctuations {\it a la} spin waves.  
The nonlinear sigma model is suitable for the description of
the 2D quantum phase transition in which the order parameter amplitude
is almost frozen, but
the strong transverse fluctuations destroy the LRO at finite temperatures.

The extension of the nonlinear sigma model approach to
the 2D quantum triangular HAF was achieved by
Dombre and Read, and Azaria et al.\cite{read,aza1}
The order parameter for this case
is expressed by a matrix with $O(3)\times O(2)$ symmetry, 
$\Phi=(\vec{\phi}_1,\vec{\phi}_2)$, where 
$\vec{\phi}_{i}^{\rm t}=(\phi_{1i},\phi_{2i},\phi_{3i})$ is an $O(3)$ vector.
The $O(3)$ symmetry reflects the spin rotational symmetry.
The $O(2)$ symmetry comes from the $E$ representation of
the $C_{3v}$ symmetry of the triangular lattice, which
is enlarged to $O(2)$ in the continuum limit.\cite{aza,del}
These vectors obey the non-linear conditions, $|\vec{\phi}_i|^2=1$, 
$\vec{\phi}_1\cdot\vec{\phi}_2=0$ corresponding to
the 120$^\circ$ structure ordered state.

Subsequently, however, it was recognized that 
the $O(3)\times O(2)$ non-linear sigma model in three dimension
is not a proper low-energy theory for 
the classical stacked triangular HAF, because the model neglects
totally longitudinal (amplitude) fluctuation modes, which
importantly induce a fluctuation-driven first order 
transition.\cite{loi,ita,tis,del} 
This fact implies that the 2D quantum version of the $O(3)\times O(2)$ 
nonlinear sigma model 
fails to capture the important low-energy physics at $T=0$.
To improve the model, one needs to introduce an effective potential which 
replaces the above nonlinear condition with the relaxed constraints.
Then, the correct low-energy effective theory for 2D quantum
triangular HAF is given by
the $O(3)\times O(2)$
matrix Ginzburg-Landau-Wilson (GLW) model,
of which the action is given by,
\begin{eqnarray}
S=S_{\sigma}+S_4+S_6, \label{model}
\end{eqnarray}
\begin{eqnarray}
S_{\sigma}=\int^{\beta}_0d\tau \int d^2x
[\frac{Z\tilde{\kappa}}{2}{\rm tr}(\partial_{\mu} \Phi^{\rm t}\partial_{\mu}\Phi)+
\frac{\tilde{\omega}\tilde{\kappa}^2}{4}V_{\alpha}\cdot V_{\alpha}],
\label{ss}
\end{eqnarray}
\begin{eqnarray}
S_4=\int^{\beta}_0d\tau \int d^2x[\frac{\tilde{\lambda}\tilde{\kappa}^2}{4}
(\frac{\hat{\rho}}{2}-1)^2+\frac{\tilde{\mu}\tilde{\kappa}^2}{4}
\hat{\tau}],
\end{eqnarray}
\begin{eqnarray}
S_6=\int^{\beta}_0d\tau \int d^2x
[\frac{\tilde{\lambda}_6\tilde{\kappa}^3}{3}
(\frac{\hat{\rho}}{2}-1)^3
+\tilde{\mu}_6\tilde{\kappa}^3(\frac{\hat{\rho}}{2}-1)\hat{\tau}],
\label{s6}
\end{eqnarray}
with $\beta=1/T$ the inverse temperature, 
and
\begin{eqnarray}
V_{a}=\epsilon_{ij}(\Phi^{\rm t}\partial_{a}\Phi)_{ij},
\end{eqnarray}
 \begin{eqnarray}
 \hat{\rho}={\rm tr}(\Phi^{\rm t}\Phi), \quad \hat{\tau}=
 \frac{1}{2}{\rm tr}(\Phi^{\rm t}\Phi-\frac{\hat{\rho}}{2})^2. 
 \end{eqnarray}
 Here, $\partial_{\mu}=(\frac{1}{c_1}\partial_{\tau},\partial_x,\partial_y)$,
 $\partial_{a}=(\frac{1}{c_3}\partial_{\tau},\partial_x,\partial_y)$.
The model (\ref{model}) describes two massless excitations
with the velocity $c_1$, and a massless excitation
with the velocity,
\begin{eqnarray}
c_t=\sqrt{\frac{Z+\tilde{\omega}\tilde{\kappa}}
{Z(c_3/c_1)^2+\tilde{\omega}\tilde{\kappa}}}c_3,
\end{eqnarray}
as well as 
a massive excitation with the mass $\tilde{\lambda}\tilde{\kappa}$, and
two massive excitations with the mass $\tilde{\mu}\tilde{\kappa}$
The three massless excitations are, respectively,
two out-of-plane modes and one in-plane 
mode of transverse fluctuations {\it a la} spin waves. 
  The second part $S_4$ is an effective potential which
 imposes the released non-linear conditions on $\vec{\phi}_i$.
 In the limit of $\tilde{\lambda}\rightarrow\infty$, 
$\tilde{\mu}\rightarrow\infty$,
 $S_4$ recovers the strict nonlinear conditions, and
$S_{\sigma}+S_4$ becomes equivalent to the action of 
the $O(3)\times O(2)$ nonlinear sigma model.\cite{hk,read,aza1,aza}
The release of the constraint allows the existence of the three
massive longitudinal fluctuation modes. 
The classical version of the model $S_{\sigma}+S_4$
has been extensively studied so far.\cite{zum,ita,tis,del}
Here we also consider 
the 6-body part $S_6$ which is required for the correct description of
 the fluctuation-driven first order transition.
In the expression of $S_6$, we neglect terms with derivatives, because 
the scaling dimensions of these terms imply that 
they are irrelevant.
Although in the following 
our analysis is applied to non-perturbative regions 
including the strong coupling limit,
we believe that 
the omission of the 6-body terms with derivatives would not change 
the essential features of our results. 
Then, since any polynomials of $\phi_i$ which preserve the $O(3) \times O(2)$
symmetry are expressed in terms of $\hat{\rho}$ and $\hat{\tau}$,\cite{del}
6-body terms are generally given by eq.(\ref{s6}).
The effective field theory (\ref{model}) captures low-energy physics of
the 2D quantum triangular HAF in the vicinity of
the long-range ordered state with the 120$^{\circ}$ structure, which
are governed by both quantum and thermal fluctuations.
The 120$^\circ$ structure state is expressed by
the configuration of $\Phi$ which minimizes $S$, i.e.
$\delta S/\delta \phi_i=0$.
Transverse and longitudinal spin fluctuations around this
configuration which preserve
relative angles between spins on a primitive triangle are included
in the model (\ref{model}). 

It is noted that although the condition $\delta S/\delta \phi_i=0$
leads the finite amplitude of the order parameter, the model
is also applicable to the disordered phase at finite temperatures,
since in 2D systems transverse fluctuations for $T> 0$ are so strong that
the average magnetization vanishes even under this condition, 
in accordance with the Mermin-Wagner-Coleman theorem.\cite{mwc}
This is a key feature of the nonlinear sigma model
as an effective field theory for the 2D quantum phase transition at $T=0$.
We would like to stress again that any perturbative approaches for
the GLW model (\ref{model}) can not describe 
the 2D quantum phase transition properly because of the reason 
explained above, and a theoretical framework which can interpolate 
the nonlinear sigma model and the GLW model is required.
The non-perturbative RG method is most suitable for this purpose,
as will be explained in the following sections.

\subsection{Non-perturbative renormalization group method for
quantum phase transitions}

We apply the non-perturbative RG method to
the model (\ref{model}).
This approach  was developed by Wetterich, Zumbach, and Delamotte et al.
for classical matrix GLW models in connection with frustrated 
magnets.\cite{wet,wet2,ber,zum2,zum,del}
A remarkable merit of this method is that in contrast to
perturbative RG calculations,
it is applicable to the whole range of the coupling constants
$\lambda$ and $\mu$ including the strong-coupling limit,
$\lambda,\mu\rightarrow\infty$.
In fact, Delamotte et al. showed that the non-perturbative RG method
successfully reproduces
the RG equations for both the 4D GLW model in the weak coupling limit and
the 2D non-linear sigma model in the strong coupling limit.\cite{del}
This feature is quite important in the investigation
of the quantum phase transition, since our systems may exhibit
the quantum-classical crossover at finite temperatures,
which is equivalent to the dimensional crossover from the 2D classical region
to the 2+1(=3)D quantum-fluctuation-dominated region. 
We exploit this fascinating advantage of the non-perturbative RG method
in the following.

It is straightforward to generalize the derivation of the RG equations
for the classical system to
the 2D quantum case.\cite{weg,wet1,pol,mor}
We assume that the effective action $\Gamma_k[\phi]$ 
for any values of a scaling parameter $k$ has the same form
as eq.(1) with the renormalized parameters; i.e. $\Gamma[\phi]=S$.
This means that the effective action is truncated up to the 6-body term.
To argue quantum critical behaviors at finite temperatures,
we utilize a non-relativistic renormalization scheme,\cite{chn,hert}
in which the infrared cutoff is introduced in the momentum space
as a scaling parameter $k$.
 The exact renormalization group equation for the quantum effective action
 $\Gamma_k[\phi]$ is given by,
\begin{eqnarray}
\partial_t\Gamma_k[\phi]=\frac{1}{2}\sum_{p,p'}\tilde{\partial}_t
\ln[\Gamma_k^{(2)}+\frac{T}{(2\pi)^2}R_k(q^2)\delta(p+p')] \label{rggam}
\end{eqnarray}
where $t=-\ln k$, $p=(i\omega_n, \vec{q})$. 
$\omega_n$ is the Matsubara frequency $2\pi n T$,  
and $\vec{q}$ is the 2D momentum.
$R_k(q^2)$ is the infrared cutoff function in the momentum space.
$\tilde{\partial}_t$ acts only on $R_k(q^2)$.
In the following, we use the theta cutoff function,
\begin{eqnarray}
R_k(q^2)=Z(k^2-q^2)\Theta(k^2-q^2).
\end{eqnarray}
$\Gamma_k^{(2)}$ is the inverse one-particle Green function defined by,
\begin{eqnarray}
\{\Gamma_k^{(2)}(p,p')\}_{\alpha\beta}=\left.\frac{\delta^2 \Gamma_k[\phi]}
{\delta\phi_{\alpha}(p)\delta\phi_{\beta}(p')}\right|_{\rm min}. 
\end{eqnarray}
Here, the derivative is taken at the configuration which
minimizes $\Gamma_k[\phi]=S$.
%

To investigate the renormalization group flow of 
the effective action (\ref{model}),
we introduce dimensionless renormalized couplings,\cite{del}
\begin{eqnarray}
\kappa=\frac{Zk^{1-d}}{c_0}\tilde{\kappa}, \quad 
\lambda=\frac{c_0\tilde{\lambda}}{Z^2k^{3-d}}, \quad
\mu=\frac{c_0\tilde{\mu}}{Z^2k^{3-d}},  \label{rc1}
\end{eqnarray}
\begin{eqnarray}
\omega=\frac{c_0\tilde{\omega}}{Z^2k^{1-d}}, \quad
\lambda_6=\frac{c_0^2}{Z^3}k^{2d-4}\tilde{\lambda}_6, \quad
\mu_6=\frac{c_0^2}{Z^3}k^{2d-4}\tilde{\mu}_6. \label{rc2}
\end{eqnarray}
Here $c_0$ is an initial value of $c_1$.
The non-perturbative RG equations for these couplings
and the velocities $c_1$ and $c_3$
are given in the appendix.
In the next sections, we present the results
obtained from numerical solutions of the RG equations.

\begin{figure}
\includegraphics*[width=8.5cm]{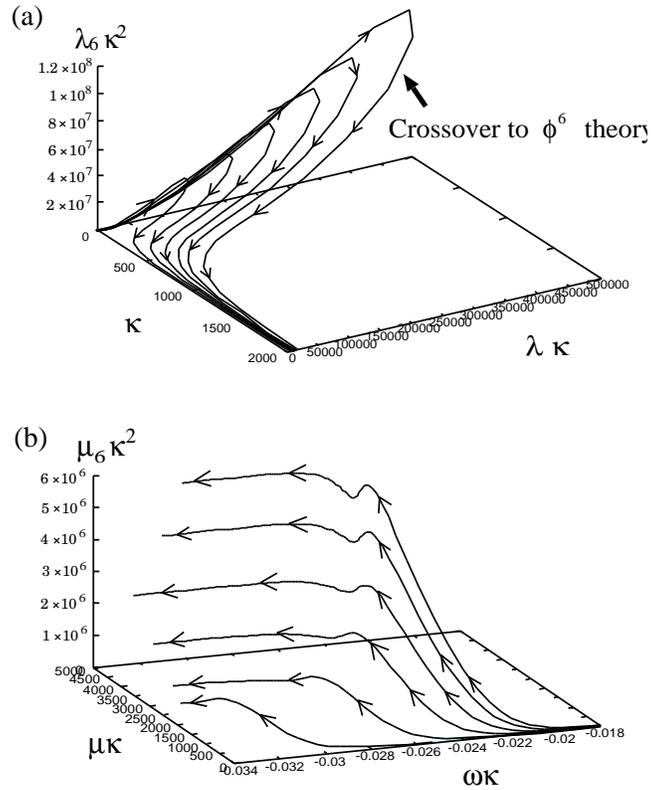}
\caption{\label{fig:fig1} RG flows of dimensionless couplings.
(a) Plot of $\kappa$, $\lambda\kappa$, and $\lambda_6\kappa^2$.
(b) Plot of $\mu\kappa$, $\omega\kappa$, and $\mu_6\kappa^2$.
Arrows indicate the directions of the RG flows.
The crossover to the $\phi^6$ theory appears in the intermediate regions}
\end{figure}

\section{Renormalization group flows for dimensionless couplings and
the phase diagram}

For the characterization of the quantum phase transition,
it is convenient to introduce a dimensionless renormalized temperature
$T'=T/(c_0k)$ with $c_0$ an initial value of $c_1$. 
At finite temperatures, in the scaling limit $k\rightarrow 0$,
the renormalized Matsubara frequency
$\omega_n=2\pi n T'$ for $n\neq 0$ becomes infinity, and thus
does not contribute to low-energy properties.
Then, in this limit, the system is in the
class of the 2D classical model, in which quantum effects are entirely included
in the renormalization of parameters.
As will be seen below, this renormalized 2D classical behaviors
appear only in the sufficiently long-wavelength scale $k\ll k_c$,
where the critical value of the scaling parameter $k_c$ is proportional
to $T$.

We solve the RG equations (\ref{kappa})-(\ref{m6}) numerically
for some particular sets of initial values of parameters
by using a Runge-Kutta-Verner method with high precision. 
We put $k=1$ at the initial stage of the renormalization.
Depending on the initial values of the parameters,
there are two regions in the scaling limit at $T=0$
as indicated in Fig.~\ref{fig:schpha}; i.e.
a long-range ordered phase, and a quantum disordered
phase (quantum paramagnet). 
The value of $1/\kappa$ at the phase boundary depends on
the choice of the initial values of the other parameters.
The correspondence between the spin $S$ triangular HAF with 
the nearest-neighbor exchange interaction $J$ and
the $O(3)\times O(2)$ non-linear sigma model implies that
the initial values of parameters are set to,\cite{read}
\begin{eqnarray}
&&c_1=\frac{3\sqrt{3}}{2\sqrt{2}}JSa, \quad 
\tilde{\kappa}=\frac{\sqrt{3}}{4}JS^2, \quad Z=1, \nonumber \\
&&\tilde{\omega}=0, \quad \frac{\tilde{\omega}}{c_3^2}=
-\frac{16}{27\sqrt{3}J^3S^4a^2}, \label{ini}
\end{eqnarray}
where $a$ is the lattice constant.
These parameters are in the region 
where the LRO exists at $T=0$, and thus we concentrate on this case
in the following.
As will be seen below, 
as long as the initial values of parameters are in this region,
the qualitative and essential features of the RG flows are not
altered by changing the parameters from those given by (\ref{ini}).

\begin{figure}
\includegraphics*[width=7cm]{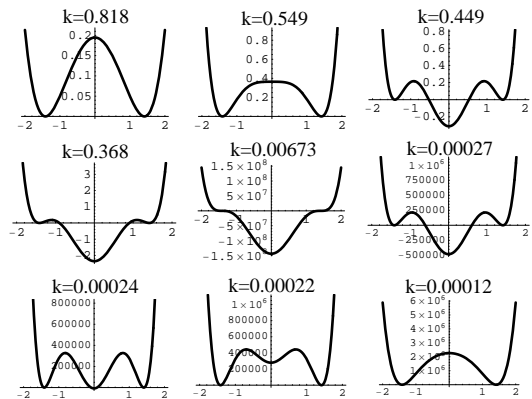}
\caption{\label{fig:fig2} Effective potentials $S_4+S_6$
as a function of $\sqrt{\hat{\rho}}$ calculated  
for a particular set of scaling parameters $k$ under the condition 
$\hat{\tau}=0$.}
\end{figure}

In Fig.~\ref{fig:fig1}, we show the RG flows obtained at $T=0.001c_0$
for some initial values of parameters.
Since the parameter $\kappa$, which is the overall coefficient of 
the action (\ref{model}), scales to infinity as $k\rightarrow 0$,
we plot the running coupling constants 
of the 4-body and 6-body terms, $\lambda\kappa^2$, $\mu\kappa^2$,
$\lambda_6\kappa^3$, and $\mu_6\kappa^3$, divided by $\kappa$ 
to specify the low-energy behaviors. 
The numerical solutions for the RG equations (\ref{kappa})-(\ref{m6})
show that 
the nature of the phase transition and long-wavelength behaviors
of the model (\ref{model})
are different from those predicted for the nonlinear sigma model.\cite{aza1}
We do not find any non-trivial fixed points which separate
the long-range ordered and disordered states,
in agreement with the recent studies on
the 3D classical stacked triangular HAF.\cite{zum,loi,ita,tis,del}
As the scaling parameter $k$ decreases, 
$\lambda \kappa$ and $\mu \kappa$ scale to large values,
implying that at finite temperatures the system is renormalized toward 
the class of the renormalized 2D classical nonlinear
sigma model.
However, it should be noted that in the intermediate region,
the 6-body fluctuations characterized by the parameters 
$\lambda_6\kappa^2$ and $\mu_6\kappa^2$ develop strongly, which may
eventually induce a fluctuation-driven first order transition
at $T=0$.

\begin{figure}
\includegraphics*[width=7cm]{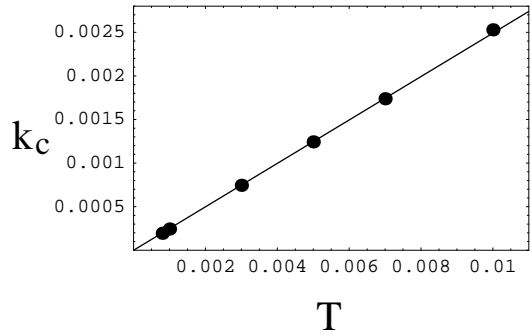}
\caption{\label{fig:fig3} Plot of the critical scaling parameter $k_c$
versus temperature $T$.}
\end{figure}

\begin{figure}
\includegraphics*[width=10cm]{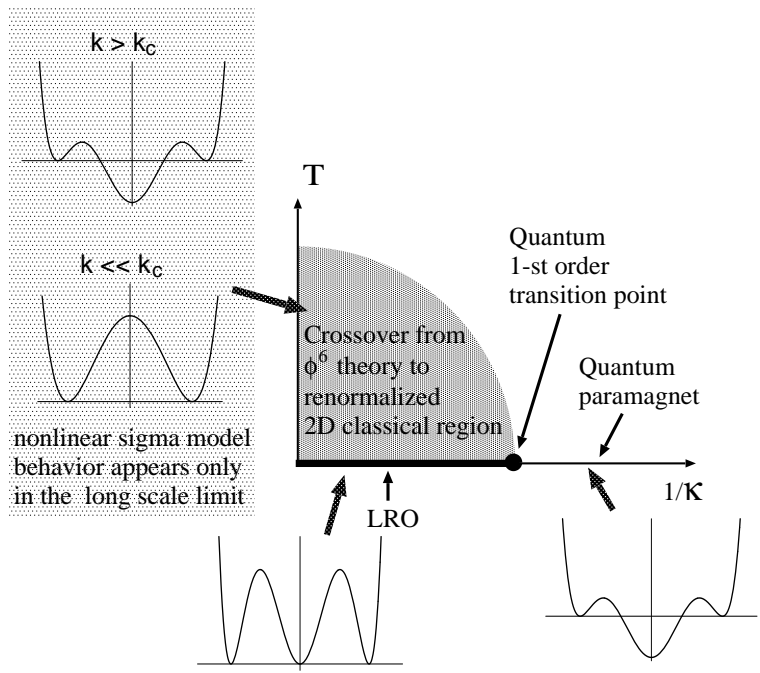}
\caption{\label{fig:fig4} A schematic phase diagram. The inserted plots
are the renormalized potential versus the renormalized field 
$\sqrt{\hat{\rho}}$. }
\end{figure}

To see the crossover behavior 
toward the $\phi^6$ model in the intermediate scale
more clearly,
we depict the renormalization of the effective potential $S_4+S_6$
for some values of the scaling parameter $k$ 
at $T=0.001c_0$ in Fig.~\ref{fig:fig2}.
Here, for simplicity, we plot the effective potential as
a function of $\sqrt{\hat{\rho}}$ under the condition
 $\hat{\tau}=0$.
In the early stage of the renormalization (k=0.818 in FIG.3), 
the effective potential
has two minima at $\sqrt{\hat{\rho}}=\pm\sqrt{2}$, corresponding to
the nonlinear condition which expresses the situation that
longitudinal fluctuations are suppressed, but strong tranverse fluctuations
inhibit the emergence of the LRO.
As $k$ decrease,
a minimum at $\sqrt{\hat{\rho}}=0$ appears, and the depth of
the valley at the origin becomes deeper and deeper,
indicating that the fluctuations which may drive the phase transition
into the first order type is developing.
At $k=0.00673$ in FIG.3, however, the growth of the minimum at the origin
stops and turns to a decline for $k<0.00673$.
Eventually, for sufficiently small $k$ ($k=0.00012$ in FIG.3), 
the minimum at the origin disappears
and the potential valleys at finite $\sqrt{\hat{\rho}}$ become deeper.
It is noted that at this final stage the effective potential 
does not describe a $\phi^4$ theory, but corresponds to
the nonlinear condition, 
showing that the system is scaled to the renormalized 2D classical 
nonlinear sigma model.
Here we introduce the scale $k_c$ which separates
the region in which $\phi^6$-type fluctuations strongly develop
($k>k_c$) and the renormalized 2D classical region($k\ll k_c$).
It is convenient to define $k_c$ as the value of $k$
for which the potential depth at the origin and
that at a finite value of $\sqrt{\hat{\rho}}$ coincide.
(In Fig.~\ref{fig:fig2}, $k_c=0.00024$.) 
At $k=k_c$, the paramagnetic state and the magnetically ordered state
are degenerate. If this situation realizes in the limit of $k\rightarrow 0$,
the first order transition to the ordered state occurs. 
Indeed, this happens at $T=0$.
In Fig.~\ref{fig:fig3}, we plot $k_c$ 
calculated for several values of temperatures $T$. The results show that
$k_c$ is proportional to $T$, and the true phase transition
which occurs in the limit $k\rightarrow 0$ 
realizes only at $T=0$ as a first order
transition, which is consistent with the results derived for 
the 3D stacked classical model.\cite{loi,ita,tis,del}
$k_c\sim T$ is also the scale at which the quantum-classical crossover
occurs.\cite{hert}
Then, as the temperature decreases toward $T=0$,
the renormalized 2D classical behaviors appear only for length 
scales much larger
than $1/k_c$, and the low-energy physics at finite temperatures
are mainly governed by the quantum fluctuations which induce
the first order transition.
In Fig.~\ref{fig:fig4}, we show the schematic phase diagram
on the $T$-$1/\kappa$ plane suggested from these RG flows
for a particular set of the initial values of the other parameters,
$\lambda$, $\mu$, etc.
The essential feature of the phase diagram depicted in Fig.~\ref{fig:fig4} 
does not depend on the choice of these initial values.
It is noted that in the long range ordered state at $T=0$,
the effective potential has degenerate minima at the origin and
at the non-zero magnetization. This is the unique feature of
the quantum first order phase transition for which the transition temperature
is $T=0$. The shape of the effective potential implies
that there may be a possible coexistence of the ordered and
disordered states.
The fluctuations which lead the crossover toward the $\phi^6$ theory
should significantly affect the low-energy properties
of the system.
We would like to address this issue in the next section.

\section{Emergent quasi-Gaussian low-energy behaviors}

Since the zero temperature phase transition in our system
is the first order type, universal critical behaviors do not exist.
However, it is still possible to relate
low-energy properties of the model (\ref{model}) appeared
in a certain parameter region to experimentally
observable quantities which are governed by quantum and thermal fluctuations.
For this purpose, we investigate the RG flows of 
dimensionful couplings $\tilde{\kappa}$, $\tilde{\lambda}$, $\tilde{\mu}$,
$\tilde{\omega}$, $\tilde{\lambda}_6$, and $\tilde{\mu}_6$
for some particular sets of initial values of the parameters, from
which physical quantities can be calculated.
Surprisingly, as we will see below, long-wavelength qualitative behaviors of
these running couplings seem to be {\it almost universal} to some extent
at least in the region where the long-range order realizes at $T=0$,
and furthermore, the effective action $\Gamma_k[\phi]$ is renormalized to
a system in which Gaussian fluctuations dominate low-energy properties.

\begin{figure}
\includegraphics*[width=8.5cm]{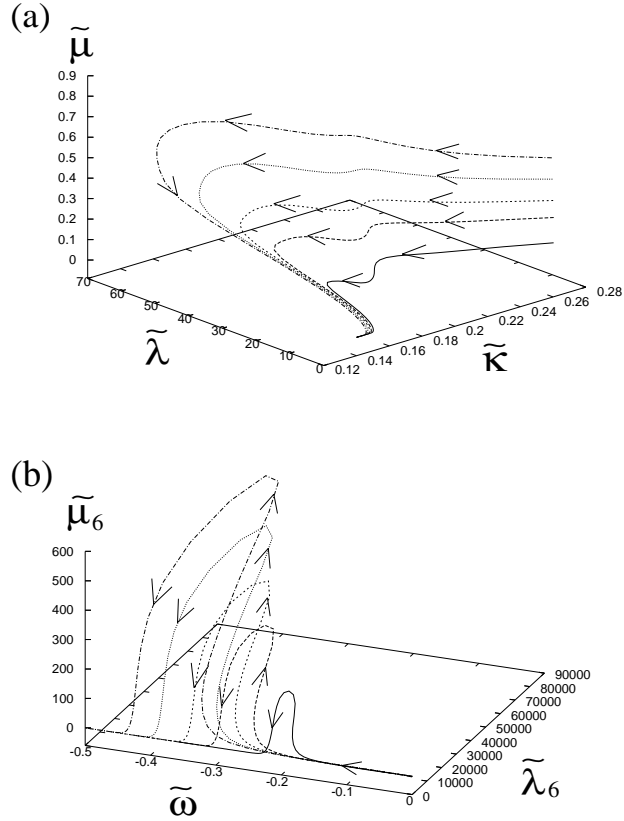}
\caption{\label{fig:fig5} RG flows of dimensionful couplings for
some particular sets of the initial values of parameters.
(a) Plot of $\tilde{\kappa}$, $\tilde{\lambda}$, and $\tilde{\mu}$.
(b) Plot of $\tilde{\omega}$, $\tilde{\lambda}_6$, and $\tilde{\mu}_6$.
The initial values of parameters are,
 $\tilde{\kappa}(t=0)=0.3$, $\tilde{\lambda}(0)=0.5$, 
$\tilde{\mu}(0)=0.1$, $\tilde{\omega}(0)=-0.0001$, 
$\tilde{\omega}(0)/c_3^2(0)=-0.5$, $c_1(0)=1.0$, (solid line);
 $\tilde{\kappa}(0)=0.3$, $\tilde{\lambda}(0)=0.3$, 
$\tilde{\mu}(0)=0.22$, $\tilde{\omega}(0)=-0.0001$, 
$\tilde{\omega}(0)/c_3^2(0)=-0.5$, $c_1(0)=1.0$, (broken line);
 $\tilde{\kappa}(0)=0.3$, $\tilde{\lambda}(0)=0.3$, 
$\tilde{\mu}(0)=0.3$, $\tilde{\omega}(0)=-0.0001$, 
$\tilde{\omega}(0)/c_3^2(0)=-0.5$, $c_1(0)=1.0$, (thin broken line);
 $\tilde{\kappa}(0)=0.3$, $\tilde{\lambda}(0)=0.3$, 
$\tilde{\mu}(0)=0.4$, $\tilde{\omega}(0)=-0.0001$, 
$\tilde{\omega}(0)/c_3^2(0)=-0.5$, $c_1(0)=1.0$, (dotted line);
 $\tilde{\kappa}(0)=0.3$, $\tilde{\lambda}(0)=0.3$, 
$\tilde{\mu}(0)=0.5$, $\tilde{\omega}(0)=-0.0001$, 
$\tilde{\omega}(0)/c_3^2(0)=-0.5$, $c_1(0)=1.0$, (dotted-and-broken line), 
and the initial values of both $\lambda_6$ and $\mu_6$ are equal to 0
in all calculations.
 }
\end{figure}

\begin{figure}
\includegraphics*[width=8.5cm]{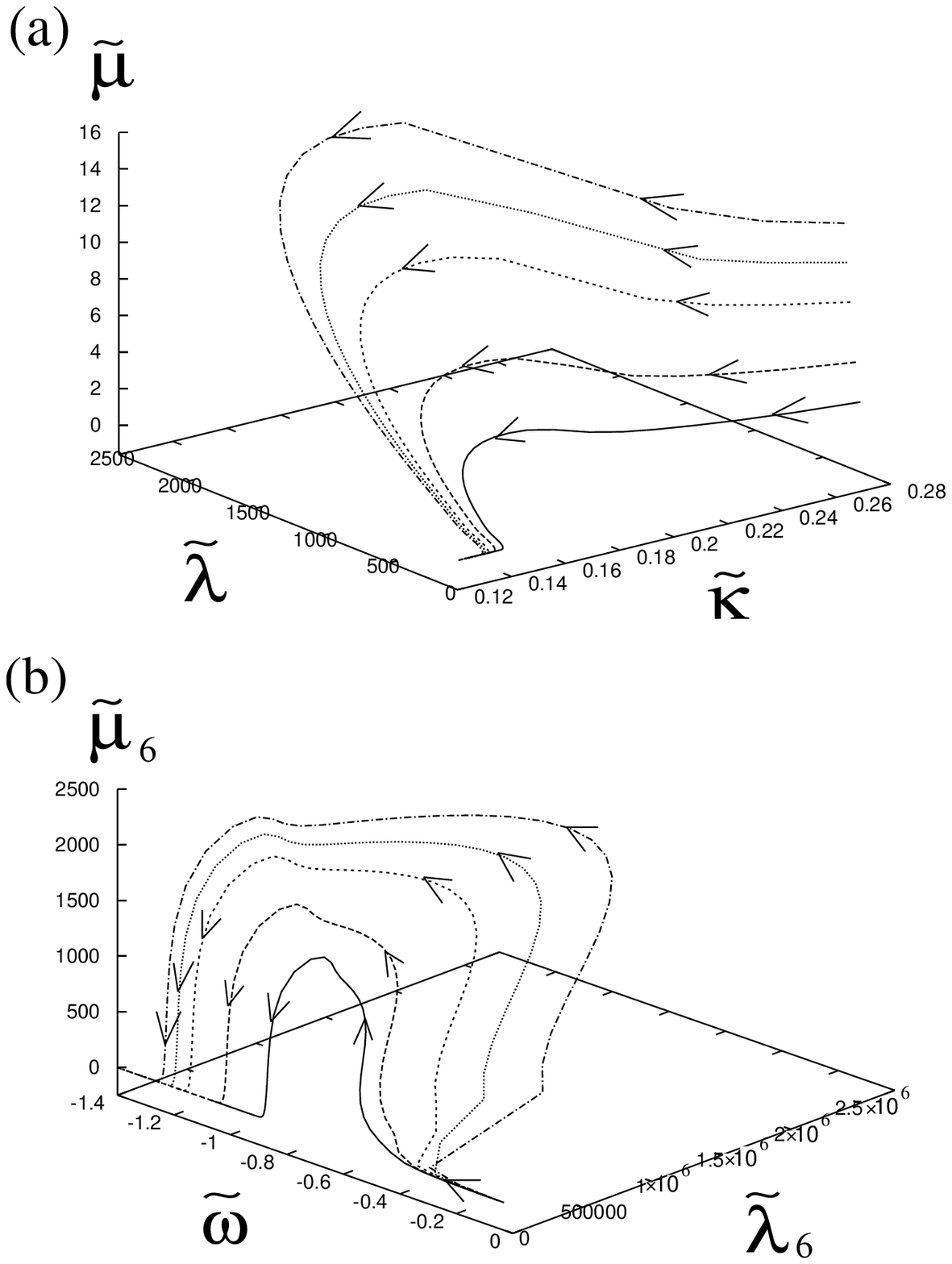}
\caption{\label{fig:fig6} RG flows of dimensionful couplings. 
The initial values of parameters are,
 $\tilde{\kappa}(t=0)=0.3$, $\tilde{\lambda}(0)=3.0$, 
$\tilde{\mu}(0)=3.0$, $\tilde{\omega}(0)=-0.0001$, 
$\tilde{\omega}(0)/c_3^2(0)=-0.5$, $c_1(0)=1.0$, (solid line);
 $\tilde{\kappa}(0)=0.3$, $\tilde{\lambda}(0)=5.0$, 
$\tilde{\mu}(0)=5.0$, $\tilde{\omega}(0)=-0.0001$, 
$\tilde{\omega}(0)/c_3^2(0)=-0.5$, $c_1(0)=1.0$, (broken line);
 $\tilde{\kappa}(0)=0.3$, $\tilde{\lambda}(0)=8.0$, 
$\tilde{\mu}(0)=8.0$, $\tilde{\omega}(0)=-0.0001$, 
$\tilde{\omega}(0)/c_3^2(0)=-0.5$, $c_1(0)=1.0$, (thin broken line);
 $\tilde{\kappa}(0)=0.3$, $\tilde{\lambda}(0)=10.0$, 
$\tilde{\mu}(0)=10.0$, $\tilde{\omega}(0)=-0.0001$, 
$\tilde{\omega}(0)/c_3^2(0)=-0.5$, $c_1(0)=1.0$, (dotted line);
 $\tilde{\kappa}(0)=0.3$, $\tilde{\lambda}(0)=12.0$, 
$\tilde{\mu}(0)=12.0$, $\tilde{\omega}(0)=-0.0001$, 
$\tilde{\omega}(0)/c_3^2(0)=-0.5$, $c_1(0)=1.0$, (dotted-and-broken line), 
and the initial values of both $\lambda_6$ and $\mu_6$ are equal to 0
in all calculations.
 }
\end{figure}

In Figs.~\ref{fig:fig5} and \ref{fig:fig6}, 
we show the RG flows for the dimensionful couplings.
It is noted that $\tilde{\lambda}$, $\tilde{\mu}$, $\tilde{\lambda}_6$,
and $\tilde{\mu}_6$ scale to zero, and $\tilde{\kappa}$ and
$\tilde{\omega}$ scale to finite non-universal constants.
Besides, $\tilde{\kappa} \gg \tilde{\kappa}^2\tilde{\omega}$
for $k\rightarrow 0$, as long as the initial value
of $\tilde{\kappa}^2\tilde{\omega}$ is much smaller than that 
of $\tilde{\kappa}$, which is a proper assumption for our system
since the $\tilde{\omega}$-term (the second term of
eq.(\ref{ss})) is generated in the process of the renormalization.
We would like to stress that these characteristic behaviors are
rather universally found for any initial values of the parameters
in the region mentioned above.
These observations imply that
in the scaling limit $k\rightarrow 0$, the effective action
is renormalized to a Gaussian-like model, which is
given by the first term of eq.(\ref{ss}) without the nonlinear conditions:
\begin{eqnarray}
S_{\rm G}=\int d^2x \int d\tau\frac{Z\tilde{\kappa}}{2}\sum_{i=1}^2
\partial_{\mu}\vec{\phi}_i\cdot\partial_{\mu}\vec{\phi}_i.
\end{eqnarray}
In the region where the Gaussian-like fluctuation, i.e.
the free boson with a linear dispersion, dominates,
the specific heat coefficient is easily calculated as,
\begin{eqnarray} 
C_v=(3\sqrt{3}/\pi)\zeta (3)(T/c_1)^2.
\end{eqnarray}
The spin susceptibility is also obtained from the Gaussian action.
For this purpose, we introduce
an external in-plane magnetic field $\vec{h}$, which couples with 
the uniform component of spin fluctuations in the form,
\begin{eqnarray}
-\frac{Z\tilde{\kappa}}{c_1^2}
\vec{h}\cdot (\vec{\phi}_1\times\partial_t\vec{\phi}_1
+\vec{\phi}_2\times\partial_t\vec{\phi}_2).
\end{eqnarray}
Then, in the quasi-Gaussian region,
the spin susceptibility in the limit $T\rightarrow 0$ 
is a non-zero constant given by,
$\chi=-\partial^2 F/\partial h^2=k_0/(12\pi c_1)$
with $k_0$ an ultra-violet momentum cutoff.
The quasi-Gaussian modes behave like spin waves with a linear dispersion,
despite the absence of spontaneous symmetry breaking at finite temperatures.
These results seem to explain partly the recent experimental
observations for the quasi-2D triangular HAF NiGa$_2$S$_4$ at low temperatures.
The spin-liquid-like behaviors found in this material at finite temperatures  
may be attributed to the existence of these fluctuation modes. 

\section{Summary and Discussion}

We have investigated
low-energy properties of the 2D quantum triangular HAF
using the non-perturbative RG method and the mapping to
the $O(3)\times O(2)$ matrix GLW model.
Our findings are as follows:

(i) At finite temperatures, at the length scale shorter than $1/k_c$
the $\phi^6$ model which describes
the fluctuations driving the transition to the first order type
dominates the low-temperature behaviors, while
at the scale larger than $1/k_c$ 
the renormalized 2D classical region appears.
As the temperature is lowered, the crossover scaling parameter $k_c$
decreases as $\propto T$, and eventually at $T=0$,
the first order phase transition to the 120$^{\circ}$ structure
occurs.

(ii) In the crossover region at finite temperatures,
the long-wavelength properties are governed by quasi-Gaussian
fluctuations {\it a la } spin waves with a linear dispersion, which
give a quadratic $T$-dependence of the specific heat coefficient and
a finite and $T$-independent non-zero 
value of the uniform spin susceptibility at low temperatures.

Here we would like to discuss the relation between our results and
the recent experimental observations 
for the quasi-2D quantum Heisenberg antiferromagnet 
on a triangular lattice, NiGa$_2$S$_4$,
which show no sign of LRO down to 37 mK, implying
the possible realization of a spin liquid.\cite{naka}
At low temperatures, the system exhibits some remarkable properties;
(a) The specific heat coefficient shows the quadratic temperature dependence
$C_v\propto T^2$, indicating the existence of gapless excitation modes. 
(b) The uniform spin susceptibility for $T\rightarrow 0$ is a nonzero
constant, suggesting that magnetic excitations are gapless. 
(c) In contrast to the observations (a) and (b), 
the magnetic correlation length measured by
the neutron scattering is rather short; i.e. $\xi \sim 2.5$ nm,
leading the authors of ref. 37 to the conclusion that there may exist
gapless non-magnetic modes.
Our results for the specific heat coefficient and the spin susceptibility
obtained in Sec.IV
seem to be in agreement with the observations (a) and (b), provided that 
the origin of the gapless excitations may be attributed to magnetic ones, 
and that in this system the coupling between the triangular layers 
is so week that thermal fluctuations suppress the true LRO
at experimentally accessible low temperatures.
However, the observation (c) implies that the magnetic excitations
may not propagate coherently and have an excitation gap, and 
is not consistent with our RG analysis 
which shows the existence of an exponentially long correlation length. 
Then, how to reconcile our results with the observation (c)?
To explain this point, we would like to note that 
in our field-theoretical model, chirality domains, which inevitably
exist in real triangular HAF and suppress the development
of the magnetic correlation length, are not included.
Also, as was pointed out by Kawamura and Miyashita\cite{kami},
$Z_2$ vortex, which is not taken into account explicitly 
in the field-theoretical model, may play a crucial role
at finite temperatures, disturbing the growth of $\xi$.\cite{z2}
It is expected that the velocities of the magnetic excitations, $c_{1,3}$, 
are strongly renormalized by these topological defects, and
reduced to small values.
Then, as long as $\xi > c/T$, the magnetic excitations behave like
gapless, and our results may be applicable.
Indeed, the experimental
observation (b) intimates the existence of gapless magnetic
excitations.
Also, to interpret the experimental observations (b) and (c) 
in a consistent way,
one might need to consider effects of randomness such as impurities,
which are beyond the scope of this paper, but may play
a crucial role in connection with the topological defects inherent in
triangular HAF. 
To confirm this speculation,
we need further studies.
We would like to address this issue in the near future.

\begin{acknowledgments}
The author is indebted to B. Delamotte, H. Kawamura, 
S. Nakatsuji, and H. Tsunetsugu
for useful discussions.
The numerical computations were partly performed at the Yukawa Institute
Computer Facility, Kyoto University.
This work was supported by a Grant-in-Aid from the Ministry of Education, 
Science, and Culture, Japan.
\end{acknowledgments}

\appendix

\section{ 
Renormalization group equations for the quantum $O(3)\times O(2)$ 
Ginzburg-Landau-Wilson model}

In this appendix, we present 
the RG equations for the dimensionless renormalized 
couplings (\ref{rc1}) and (\ref{rc2}) for 
the effective action (\ref{model}) as functions of
the scaling parameter $t=-\ln k$.
Here we consider the general case of the spatial dimension $d$.
Differentiating 
eq.(\ref{rggam}) with respect to $\phi_i$, as was done in ref. 46,
we end up with the RG equations truncated up to the 6-body terms, 
\begin{eqnarray}
\partial_t \kappa&=&-(d-1+\eta)\kappa-\frac{3}{2}l_1(\kappa\lambda)
-l_1(\kappa\mu)-\frac{r_{10}}{2}-(N-2)l_1(0)-\frac{\omega}{\lambda}r_{11}
-2\frac{\mu}{\lambda}l_1(\kappa\mu) \nonumber \\
&&-\frac{4\kappa}{\lambda}[
\lambda_6l_1(\kappa\lambda)+2\mu_6l_1(\kappa\mu)], \label{kappa}
\end{eqnarray}
\begin{eqnarray}
\partial_t\lambda&=&(d-3+2\eta)\lambda-\frac{\lambda^2}{4}
(9l_2(\kappa\lambda)+2l_2(\kappa\mu)+r_{20}+2(N-2)l_2(0)) \nonumber \\
&&-2\lambda\mu l_2(\kappa\mu)-\lambda\omega r_{21}-\omega^2r_{22}
-2\mu^2l_2(\kappa\mu) \nonumber \\
&&+4\lambda_6l_1(\kappa\lambda)+68\mu_6l_1(\kappa\mu)
-4\frac{\lambda_6}{\lambda}[\omega r_{11}+2\mu l_1(\kappa\mu)] \nonumber \\
&&-16\frac{\kappa}{\lambda}[\lambda_6^2 l_1(\kappa\lambda)+2\lambda_6\mu_6 
l_1(\kappa\mu)]-16\lambda_6^2\kappa^2 l_2(\kappa\lambda)
-32\mu_6^2\kappa^2 l_2(\kappa\mu)   \nonumber \\
&&-12\kappa\lambda\lambda_6 l_2(\kappa\lambda)
-8\kappa(\lambda+2\mu)\mu_6 l_2(\kappa\mu), \label{lambda}
\end{eqnarray}
\begin{eqnarray}
\partial_t \mu&=&(d-3+2\eta)\mu-3\mu\lambda u_{11}(\kappa\lambda,\kappa\mu) 
-\mu\omega r_{111}(\kappa\mu)  \nonumber \\
&-&\frac{\mu^2}{2}
[3u_{11}(\kappa\lambda,\kappa\mu)+(N-2)l_2(0)-r_{110}(\kappa\mu)] \nonumber \\
&+&4\mu_6[l_1(\kappa\lambda)+l_1(\kappa\mu)-\frac{\omega}{\lambda}r_{11}
-\frac{2\mu}{\lambda}l_1(\kappa\mu)]
-\frac{16\kappa}{\lambda}[\lambda_6\mu_6+2\mu_6^2 l_1(\kappa\mu)] \nonumber \\
&-&32\kappa^2\mu_6^2 u_{11}(\kappa\lambda,\kappa\mu)-8\kappa\mu_6
(\lambda+2\mu)u_{11}(\kappa\lambda,\kappa\mu), \label{mu}
\end{eqnarray}
\begin{eqnarray}
\eta&=&-\frac{\omega}{2}l_1(0)-\frac{\kappa\lambda^2}{2}l_{120}(\kappa\lambda)
+\kappa\omega^2[2h_{111}(0)+h_{110}(0)+(1+\kappa\omega)h_{222}(0)
+2r_{11\tilde{1}}(0)] \nonumber \\
&&-\kappa\mu^2 l_{120}(\kappa\mu)  \label{eta}
\end{eqnarray}
\begin{eqnarray}
\partial_t c_1&=&\frac{c_1\omega}{4}\biggl[1-\frac{c_1^2}{c_3^2}\biggr]
l_1(0) \nonumber \\
&+&\frac{c_1\kappa\lambda^2}{4}[-3l_{220}(\kappa\lambda)+
2l_{210}(\kappa\lambda)+4l_{300}(\kappa\lambda)-4l_{310}(\kappa\lambda)] \nonumber \\
&+&\frac{c_1\kappa\mu^2}{4}[-3l_{220}(\kappa\mu)+
2l_{210}(\kappa\mu)+4l_{300}(\kappa\mu)-4l_{310}(\kappa\mu)] \nonumber \\
&+&\frac{c_1\kappa\omega^2}{4}[4\frac{c_1^2}{c_3^2}r_{11\check{1}}(0)
-r_{121}(0)+4r_{12\tilde{2}}(0)-4r_{12\check{2}}(0)-4r_{132}(0)+4w_{132}(0) 
\nonumber \\
&-&4h_{111}(0)-2h_{111}(0)-2(1+\kappa\omega)h_{222}(0)-2r_{11\tilde{1}}(0)],
\label{c1}
\end{eqnarray}
\begin{eqnarray}
\partial_t \omega &=&(d-1+2\eta)\omega+\frac{\omega}{2\kappa}
[l_1(\kappa\lambda)-l_1(0)]-\frac{\lambda^2}{2}
[l_{120}(\kappa\lambda)+2(1+\kappa\omega)h_{220}(\kappa\lambda)] \nonumber \\
&+&\omega^2[2h_{111}(0)+h_{110}(0)+(1+\kappa\omega)h_{222}(0)+r_{11\tilde{1}}(0)
\nonumber \\
&-&6h_{121}(\kappa\lambda)-(1+\kappa\omega)h_{222}(\kappa\lambda)
-r_{111}(\kappa\lambda)-\frac{8}{d}r_{11\tilde{1}}(\kappa\lambda) \nonumber \\
&+&\frac{8}{d+2}h_{030}(\mu\kappa)+\frac{8}{d+2}(N-2)h_{030}(0)]
-\mu^2l_{120}(\kappa\mu) \nonumber \\
&+&\lambda\omega[6h_{120}(\kappa\lambda)
+(1+\kappa\omega)h_{221}(\kappa\lambda)+r_{110}(\kappa\lambda)], \label{omega}
\end{eqnarray}
\begin{eqnarray}
\partial_t c_3&=&\frac{c_3\lambda^2}{4\omega}
[l_{120}(\kappa\lambda)+2(1+\kappa\omega)h_{220}(\kappa\lambda)
-\frac{c_3^2}{c_1^2}\{-l_{120}(\kappa\lambda)-3l_{220}(\kappa\lambda)+
2l_{210}(\kappa\lambda) \nonumber \\
&+&4l_{300}(\kappa\lambda) 
-4l_{310}(\kappa\lambda)
-3w_{220}(\kappa\lambda)-10w_{130}(\kappa\lambda) \nonumber \\ 
&+&4(1+\kappa\lambda)w_{230}(\kappa\lambda) 
+12(1+\kappa\lambda)w_{140}(\kappa\lambda) \}]  \nonumber \\ 
&+&\frac{c_3\omega}{2}[2h_{111}(0)+h_{110}(0)+(1+\kappa\omega)h_{222}(0)
+r_{11\tilde{1}}(0)
\nonumber \\
&-&6h_{121}(\kappa\lambda)-(1+\kappa\omega)h_{222}(\kappa\lambda)
-r_{111}(\kappa\lambda)-\frac{8}{d}r_{11\tilde{1}}(\kappa\lambda)\nonumber \\
&+&\frac{8}{d+2}h_{030}(\mu\kappa)+\frac{8}{d+2}(N-2)h_{030}(0)  \nonumber \\
&-&\frac{1}{2}\frac{c_3^2}{c_1^2}\{4\frac{c_1^2}{c_3^2}r_{11\check{1}}(0)
-r_{121}(0)+4r_{12\tilde{2}}(0) 
-4r_{12\check{2}}(0)-4r_{132}(0)  \nonumber \\
&+&4w_{132}(0)  +6r_{111}(\kappa\lambda)  + 8w_{121}(\kappa\lambda)
-16r_{11\check{1}}(\kappa\lambda)-16(1+\kappa\lambda)r_{121}(\kappa\lambda)
\nonumber \\
&+&16h_{03\check{1}}(\mu\kappa)+16(N-2)h_{03\check{1}}(0)\}]  \nonumber \\
&+&\frac{c_3\mu^2}{2\omega}[-l_{120}(\kappa\lambda)-\frac{c_3^2}{c_1^2}
(-l_{120}(\kappa\lambda)-3l_{220}(\kappa\lambda)  \nonumber \\
&+&2l_{210}(\kappa\lambda)+4l_{300}(\kappa\lambda)-4l_{310}(\kappa\lambda))]
\nonumber \\
&-&\frac{c_3\lambda}{2}[-6h_{120}(\kappa\lambda)
-(1+\kappa\omega)h_{221}(\kappa\lambda)-r{110}(\kappa\lambda) \nonumber \\
&-&\frac{c_3^2}{c_1^2}\{-3r_{121}(\kappa\lambda)+4r_{12\check{1}}(\kappa\lambda)
-4w_{131}(\kappa\lambda)+4r_{131}(\kappa\lambda)
\}], \label{c3}
\end{eqnarray}
\begin{eqnarray}
\partial_t \lambda_6&=&(2d-4+3\eta)\lambda_6+
16 \kappa^2 \lambda_6^3 l_3(\kappa\lambda) 
-30 \kappa\lambda_6^2
   l_2(\kappa\lambda) 
-12 \kappa \mu_6\lambda_6 l_2(\kappa\mu) \nonumber \\ 
&-&3 \mu  \lambda_6 l_2(\kappa\mu) 
-\frac{3 \omega   
\lambda_6}{2}r_{21} 
+32 \kappa^2\mu_6^3 l_3(\kappa\mu)
+\frac{\mu ^3}{2 \kappa}l_3(\kappa\mu) 
-48\kappa \mu_6^2l_2(\kappa\mu)  \nonumber \\
&+&6 \mu_6\mu ^2l_3(\kappa\mu) 
-\frac{39 \mu_6}{2 \kappa}l_1(\kappa\mu) +24
   \kappa \mu_6^2 \mu l_3(\kappa\mu) 
-12 \mu_6 \mu l_2(\kappa\mu)   \nonumber \\
&+&\frac{\lambda^3}{32\kappa} (2(N-2)l_3(0)
   +27 l_3(\kappa\lambda)+2l_3(\kappa\mu)+r_{30})    \nonumber \\
&+& \frac{27\lambda^2 
   \lambda_6}{4}l_3(\kappa\lambda)+\frac{3 \mu_6\lambda^2}{2}l_3(\kappa\mu) 
+\frac{3\mu\lambda^2}{8 \kappa}l_3(\kappa\mu)  +\frac{3 \omega\lambda^2    
}{16 \kappa}r_{31}   \nonumber \\
&+&
18 \kappa \lambda \lambda_6^2  l_3(\kappa\lambda) 
-\frac{\lambda \lambda_6}{4}(6(N-2)l_2(0)+45 l_2(\kappa\lambda) 
+ 6 l_2(\kappa\mu) 
+ 3 r_{20}) \nonumber \\
&+&12 \kappa \lambda \mu_6^2l_3(\kappa\mu) 
+\frac{3\lambda \mu ^2}{4 \kappa}   l_3(\kappa\mu) 
-6\lambda \mu_6 l_2(\kappa\mu) \nonumber \\
&+&6\lambda \mu_6 \mu  l_3(\kappa\mu) 
+\frac{3 \lambda \omega^2
   r_{32}}{8 \kappa}
+\frac{\omega^3 r_{33}}{4 \kappa}, \label{l6}
\end{eqnarray}
\begin{eqnarray}
\partial_t \mu_6&=&(2d-4+3\eta)\mu_6+
\frac{\lambda^3}{16} (3
   u_{21}(\kappa\lambda,\kappa\mu)+u_{21}(\kappa\mu,\kappa\lambda))
  \nonumber \\
&+&\frac{1}{8} (-\frac{u_{11}(\kappa\lambda,\kappa\mu)}{\kappa }+\frac{3
   l_2(\kappa\lambda)}{2 \kappa }-\frac{l_2(\kappa\mu)}{2 \kappa }  \nonumber \\
&+&4\kappa  u_{21}(\kappa\lambda,\kappa\mu) \lambda_6
+24 \kappa  u_{21}(\kappa\lambda,\kappa\mu)
   \mu_6+12 \kappa  u_{21}(\kappa\mu,\kappa\lambda) \mu_6  \nonumber \\
&+&6 u_{21}(\kappa\lambda,\kappa\mu)
   \mu +3 u_{21}(\kappa\mu,\kappa\lambda) \mu ) \lambda ^2 \nonumber \\
&+&\frac{1}{8}
   (96 u_{21}(\kappa\lambda,\kappa\mu) \mu_6^2 \kappa ^2
+96 u_{21}(\kappa\mu,\kappa\lambda)
   \mu_6^2 \kappa ^2+64 u_{21}(\kappa\lambda,\kappa\mu) \lambda_6 \mu_6
   \kappa ^2 \nonumber \\
&+&16 u_{21}(\kappa\lambda,\kappa\mu) \lambda_6 \mu  \kappa +48
   u_{21}(\kappa\lambda,\kappa\mu) \mu_6 \mu  \kappa  \nonumber \\ 
&+&48 u_{21}(\kappa\mu,\kappa\lambda)
   \mu_6 \mu  \kappa +6 u_{21}(\kappa\lambda,\kappa\mu) \mu ^2
+6 u_{21}(\kappa\mu,\kappa\lambda)
   \mu ^2+l_3(0) \mu ^2   \nonumber \\
&+&4 l_2(\kappa\lambda) \lambda_6-4
   l_2(0) \mu_6-30 l_2(\kappa\lambda) \mu_6-24
   l_2(\kappa\mu) \mu_6 \nonumber  \\
&-&8 u_{11}(\kappa\lambda,\kappa\mu) (\lambda_6+4
   \mu_6)-2 \mu_6 r_{20}+2 \omega ^2
   r_{212}(\kappa\mu)+\frac{2 u_{11}(\kappa\lambda,\kappa\mu) \mu }{\kappa }
\nonumber \\
   &-&\frac{3
   l_2(\kappa\lambda) \mu }{2 \kappa }-\frac{l_{201}(\kappa\mu) \omega }{2
   \kappa }-\frac{\mu  r_{20}}{2 \kappa }+\frac{\omega 
   r_{21}}{2 \kappa }) \lambda  \nonumber \\
&+&\frac{1}{8}
   (256 \mu_6^2 (u_{21}(\kappa\lambda,\kappa\mu) \lambda_6
+u_{21}(\kappa\mu,\kappa\lambda)
   \mu_6) \kappa ^3   \nonumber \\
&+&192 u_{21}(\kappa\mu,\kappa\lambda) \mu_6^2 \mu 
   \kappa ^2+128 u_{21}(\kappa\lambda,\kappa\mu) \lambda_6 \mu_6 \mu  \kappa
   ^2  \nonumber \\
&+&16 u_{21}(\kappa\lambda,\kappa\mu) \lambda_6 \mu ^2 \kappa 
+48 u_{21}(\kappa\mu,\kappa\lambda)
   \mu_6 \mu ^2 \kappa -16 \mu_6 (4 u_{11}(\kappa\lambda,\kappa\mu)
   \lambda_6 \nonumber \\
&+&5 l_2(\kappa\lambda) \lambda_6+16 u_{11}(\kappa\lambda,\kappa\mu)
   \mu_6+10 l_2(\kappa\mu) \mu_6) \kappa  \nonumber \\
&+&4 u_{21}(\kappa\mu,\kappa\lambda)
   \mu ^3-4 l_2(\kappa\lambda) \lambda_6 \mu -8 l_2(0) \mu_6
   \mu -32 l_2(\kappa\mu) \mu_6 \mu  \nonumber \\
&-&16 u_{11}(\kappa\lambda,\kappa\mu)
   (\lambda_6+4 \mu_6) \mu -4 l_{201}(\kappa\mu) \mu_6
   \omega -4 \mu_6 \omega  r_{21} \nonumber \\
&+&4 \omega ^3
   r_{213}(\kappa\mu)
-\frac{u_{11}(\kappa\lambda,\kappa\mu) \mu ^2}{\kappa }+\frac{2
   l_2(\kappa\mu) \mu ^2}{\kappa } \nonumber \\
&+&\frac{4 l_1(\kappa\mu)
   \lambda_6}{\kappa }-\frac{4 l_2(\kappa\lambda) \lambda_6}{\kappa
   }-\frac{44 l_1(\kappa\mu) \mu_6}{\kappa
   }-\frac{l_{201}(\kappa\mu) \mu  \omega }{\kappa }-\frac{\mu ^2
   r_{110}(\kappa\mu)}{\kappa } \nonumber \\
&+&\frac{2 \mu  \omega 
   r_{111}(\kappa\mu)}{\kappa }-\frac{\mu  \omega 
   r_{21}}{\kappa }-\frac{\omega ^2 r_{112}(\kappa\mu)}{\kappa
   }+\frac{\omega ^2 r_{22}}{\kappa }). \label{m6}
\end{eqnarray}
Here, $\eta=-\partial_t \ln Z$ is the anomalous dimension, and
the threshold functions appeared in the above expressions are,
\begin{eqnarray}
l_{mns}(X)=\frac{v_d}{d}T'
\sum_{i}\frac{-2m(\frac{\varepsilon^2_i}{c_3^2}+\frac{d}{d+2})^s}
{(\frac{\varepsilon^2_i}{c_1^2}+1+X)^{m+1}(\frac{\varepsilon^2_i}{c_1^2}+1)^n},
\end{eqnarray}
\begin{eqnarray}
r_{mn}=v_d\int^1_0 dyy^{d-1}T'\sum_i\frac{-2m(\frac{\varepsilon^2_i}{c_3^2}
+y^2)^n}{(\frac{\varepsilon^2_i}{c_1^2}+1
+\omega\kappa(\frac{\varepsilon^2_i}{c_3^2}
+y^2))^{m+1}},
\end{eqnarray}
\begin{eqnarray}
u_{mn}(X,Y)=\frac{v_d}{d}T'\sum_i\frac{-2[(m+n)(\frac{\varepsilon^2_i}{c_1^2}+1)
+mY+nX]}{(\frac{\varepsilon^2_i}{c_1^2}+1+X)^{m+1}
(\frac{\varepsilon^2_i}{c_1^2}+1+Y)^{n+1}},
\end{eqnarray}
\begin{eqnarray}
w_{mns}(X)=v_d\int^1_0 dyy^{d-1}T'\sum_i\frac{-2(\frac{\varepsilon^2_i}{c_3^2}
+y^2)^s}
{(\frac{\varepsilon^2_i}{c_1^2}+1+\omega\kappa(\frac{\varepsilon^2_i}{c_3^2}
+y^2))^m(\frac{\varepsilon^2_i}{c_1^2}+1+X)^n},
\end{eqnarray}
\begin{eqnarray}
h_{mns}(X)=\frac{v_d}{d}T'\sum_i\frac{(\frac{\varepsilon^2_i}{c_3^2}
+y^2)^s}
{(\frac{\varepsilon^2_i}{c_1^2}+1+\omega\kappa(\frac{\varepsilon^2_i}{c_3^2}
+y^2))^m(\frac{\varepsilon^2_i}{c_1^2}+1+X)^n},
\end{eqnarray}
\begin{eqnarray}
r_{mns}(X)=v_d\int^1_0 dyy^{d-1}T'\sum_i\frac{-2[(m+n)
(\frac{\varepsilon^2_i}{c_1^2}+1)+mX+n\omega\kappa
(\frac{\varepsilon^2_i}{c_3^2}+y^2)]}
{(\frac{\varepsilon^2_i}{c_1^2}+1+\omega\kappa(\frac{\varepsilon^2_i}{c_3^2}
+y^2))^{m+1}(\frac{\varepsilon^2_i}{c_1^2}+1+X)^{n+1}}\Xi^s,
\label{rmns}
\end{eqnarray}
with $v_d=d\pi^{d/2}/[(2\pi)^d\Gamma(d/2+1)]$, $\varepsilon_n=2\pi n T'$, 
and $\Xi=\frac{\varepsilon^2_i}{c_3^2}+y^2$.
$r_{mn\tilde{s}}(X)$ is given by (\ref{rmns}) with $\Xi$ replaced by $y^2$.
$r_{mn\check{s}}(X)$ is given by (\ref{rmns}) with $\Xi$ replaced by 
$\varepsilon^2_i/c_3^2$.
Also, $l_m(X)\equiv l_{m00}(X)$.
The renormalized temperature $T'$ obeys
the RG equation $\partial_t T'=-T'$.
In the absence of the 6-body terms, 
and in the limit $T\rightarrow 0$, 
the above RG equations agree with those
obtained by Delamotte et al. for the classical model truncated
up to the forth order terms.\cite{del}
The RG equations are non-perturbative in the sense that
the beta functions (the right-hand sides of eqs.(\ref{kappa})-(\ref{m6})) 
do not explode even in the strong coupling limit 
$\lambda\rightarrow\infty$, $\mu\rightarrow\infty$.
The one-loop results of the nonlinear sigma model are reproduced
in this limit.


\begin{thebibliography}{99}
\bibitem{z2} The role of the $Z_2$ vortex was, originally, discussed
by Kawamura and Miyashita\cite{kami} in connection with a possibility of
KT-like phase transition in 2D HAF. As a matter of fact, 
KT-like low temperature critical phases caused by
binding of $Z_2$ vortices do not exist, because
in 2D systems the presence of continuous critical lines is allowed only 
in the case with $U(1)$ symmetry which
is the class of the $c=1$ conformal field theory.
\end{thebibliography}

\end{document}